\documentclass[multi]{cambridge6A}
\usepackage[numbers]{natbib}
\usepackage{rotating}
\usepackage{floatpag,ubec}
\rotfloatpagestyle{empty}
\usepackage{amsthm,amsmath}
\usepackage{graphicx}
\usepackage[usenames,dvipsnames]{xcolor}

\def\gappeq{\mathrel{ \rlap{\raise.5ex\hbox{$>$}}
                      {\lower.5ex\hbox{$\sim$}} } }
\def\lappeq{\mathrel{ \rlap{\raise.5ex\hbox{$<$}}
                      {\lower.5ex\hbox{$\sim$}} } }

\begin{document}

%

\author{N. G. Parker, A. J. Allen, C. F. Barenghi and N. P. Proukakis}
\affiliation{Joint Quantum Centre (JQC) Durham-Newcastle, School of Mathematics and Statistics, Newcastle University, Newcastle upon Tyne, United Kingdom}
\chapter{A quantum storm in a teacup}

\begin{abstract}
The past decade has seen atomic Bose-Einstein condensates emerge as a promising prototype system to explore the quantum mechanical form of turbulence, buoyed by a powerful experimental toolbox to control and manipulate the fluid, and the amenity to describe the system from first-principles.  This article presents an overview of quantum turbulence in atomic condensates, from its history and fundamental motivations, its characteristics and key results to date, and finally to some promising future directions. 
\end{abstract}

\section{Introduction}

\begin{figure}
\includegraphics[width=0.4\textwidth]{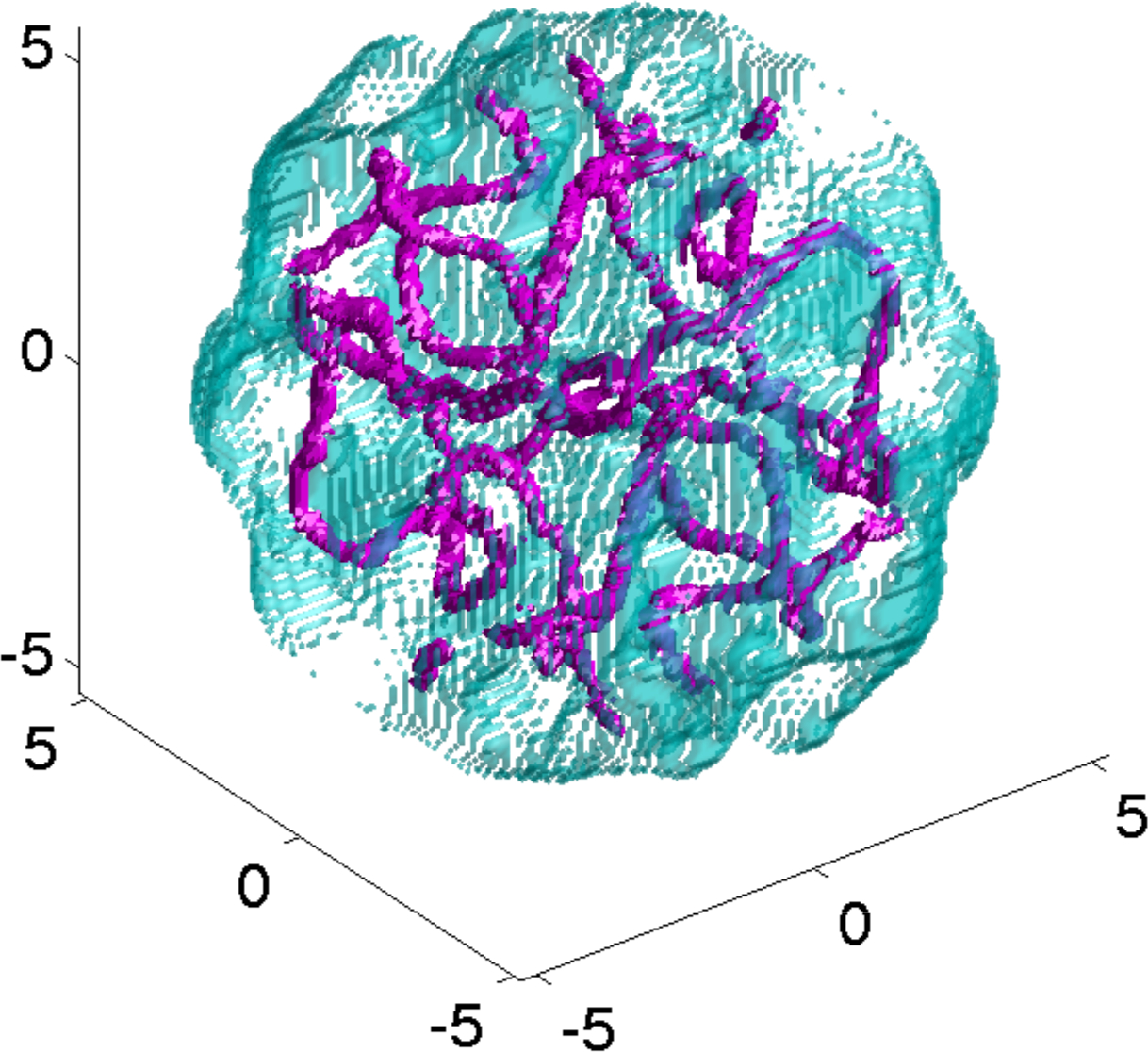} \hfill
\includegraphics[width=0.4\textwidth]{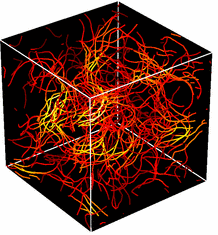} 
\caption{Left: A turbulent state in a trapped atomic Bose-Einstein condensate, highlighting the tangle of vortex cores (purple/dark grey) and the condensate surface (cyan/light grey).  Figure taken from Ref. \cite{white_2010}.   Right:  A vortex tangle in a large homogeneous box, applicable to turbulent Helium-II.  Vortex lines are colored according to the magnitude of the coarse-grained vorticity, with light (dark) regions corresponding to high (low) vorticity.  Light regions correspond to bundles of parallel vortex lines.  Figure taken from Ref. \cite{baggaley_2012b}. Both images correspond to numerical simulations.}
\label{fig:parker_fig1}
\end{figure}

``A quantum storm in a teacup" provides an apt metaphor for the topic of this article.    The storm refers to a turbulent state of a fluid, teeming with swirls and waves.  Quantum refers to the fact that the fluid is not the classical viscous fluid of conventional storms but rather a quantum fluid in which viscosity is absent and the swirls are quantized.   The quantum fluid in our story is a quantum-degenerate gas of bosonic atoms, an atomic Bose-Einstein condensate (BEC), formed at less than a millionth of a degree above absolute zero.  And finally the teacup refers to the bowl-like potential used to confine the gas; this makes the fluid inherently inhomogeneous and finite-sized.  A typical image of our quantum storm in a teacup is shown in Fig. \ref{fig:parker_fig1}(a).  

This chapter reviews quantum turbulence in atomic condensates, tracing its history (Section \ref{sec:history}) and introducing the main theoretical approach (Section \ref{sec:theory}) and the underyling quantum vortices (Section \ref{sec:vortices}).  We then turn to describing physical characteristics (Section \ref{sec:nature}), the experimental observations to date (Section \ref{sec:experiments}), methods of generating turbulence (Section \ref{sec:generation}) and some exciting research directions (Section \ref{sec:exotica}), before presenting an outlook (Section \ref{sec:conclusions}).

\section{Origins}
\label{sec:history}
Turbulence refers to a highly-agitated, disordered and nonlinear fluid motion, characterised by the presence of eddies and energy across a range of length and time scales  \cite{frisch_1995}.  It occurs ubiquitously in nature, from blood flow and waterways to atmospheres and the interstellar medium, and is of practical importance in many industrial and engineering contexts.  Since da Vinci's first scientific study of turbulent flow of water past obstacles, circa 1507, research into turbulence in classical viscous fluids continues with vigour; however, due to its rich complexities,  the physical essense and mathematical description of turbulence remains a challenge.  

The seeds for quantum turbulence were set by Feynman in 1955 \cite{feynman_1955} for the only known quantum fluid at that time, Helium II (bosonic $^4$He in its superfluid phase below $2.2$K).  Working from Onsager's proposal that superfluid vortices carried quantized circulation\footnote{This seminal proposal was made as a comment to a paper by Gorter \cite{gorter_1949}; see comments therein.}, he pictured the ensuing ``quantum turbulence''\footnote{The term  ``quantum turbulence" was not coined until some 30 years later by Donnelly and Swanson \cite{donnelly_1986}.} as a disordered tangle of quantized vortex lines and vortex rings.  A modern incarnation of Feynman's concept is depicted in Fig. \ref{fig:parker_fig1}(b).  Feynman may have seen quantum fluids as a prototypical perspective to approach classical turbulence, which he acknowledged as ``the most important unsolved problem of classical physics" \cite{feynman_1964}.  Soon after, Vinen pioneered the experimental study of turbulence in Helium II \cite{vinen_1957a}, driven by thermally-induced counterflow (whereby the normal and superfluid components are driven to flow in opposing directions).  With no classical analog to thermal counter-flow, these studies remained disparate from classical turbulence.  However, when experiments in the 1990s turned to conventional methods of generating turbulence, e.g. moving grids and counter-rotating discs,  striking similarities began to emerge between quantum and classical turbulence, most notably the appearance of energy spectra which follow the classical Kolmogorov behaviour \cite{maurer_1998,salort_2010}.  The current status of quantum turbulence in Helium can be found in recent reviews \cite{skrbek_2011,barenghi_2014}.  For example, questions are being asked about the role played in turbulence by vortex reconnections \cite{kerr_2011,baggaley_2014}, and about a new, non-classical quantum turbulent regime (called ``Vinen" or ``ultraquantum" turbulence) which, according to experiments \cite{walmsley_2008} and numerical simulations \cite{baggaley_2012} is different from ordinary turbulence in terms of energy spectrum and decay. Hopefully new methods of flow visualization will clarify this problem \cite{guo_2010}. In the very low temperature limit of a pure superfluid, experiments are concerned with quantum turbulence in $^3$He-B \cite{bradley_2011,hosio_2013}, a fermionic superfluid, and with a new, non-classical turbulent energy cascade \cite{krstulovic_2012}, similar to the Kolmogorov cascade, but arising from interacting Kelvin waves along vortices at very large wave numbers \cite{kozik_2004}.

The realization of gaseous atomic Bose-Einstein condensates in 1995 \cite{anderson_1995,davis_1995} introduced a new and meritable system in which to analyse the dynamics of quantum fluids and vortices therein.  Due to their weak interactions, condensate fractions of over 95\% are achievable in the limit of zero temperature (which in practice corresponds to $T \ll  T_{\rm C}$, where $T_{\rm C}$ is the critical temperature for condensatation).  A practical implication of this is that the atomic density, readily imaged via absorption and phase contrast techniques, provides direct visualization of the condensate and vortices.  Techniques from atomic physics enable vast control over the physical nature of the system.  For example, the strength of the atomic interactions can be precisely tuned (by use of a molecular Feshbach resonance \cite{inouye_1998}), the potential $V({\bf r},t)$ used to confine the condensates can be varied almost arbitrarily in both time and space \cite{henderson_2009}, condensates of reduced dimensionality can be formed \cite{gorlitz_2001}, and additional condensate species can be introduced to create multi-component quantum fluids \cite{kevrekidis_frantzeskakis_book_08}.  

In the past decade atomic BECs have emerged as a promising new setting for studying quantum turbulence \cite{tsubota_2013,allen_2014_review,white_2014}.  This began with theoretical proposals in the mid 2000's \cite{berloff_2002,parker_2005,kobayashi_2007,tsubota_2008}, with recent landmark experimental reports of quantum turbulence in three \cite{henn_2009} and two-dimensions \cite{neely_2010,kwon_2014}.  The key motivation is their embodiment of a highly controllable, prototype system in which to elucidate quantum turbulence (and many-vortex dynamics in general) from the bottom up.  However, as we will see, fundamental questions exist over the extent to which turbulence can be generated and observed in such small and inhomogeneous systems.

\section{Theory}
\label{sec:theory}
An attractive feature of atomic condensates is their amenity to first-principles theoretical modelling.  Due to their high condensate fraction, diluteness and weak interactions, the system is well described by the microscopic theory of the weakly-interacting Bose gas \cite{pethick_smith_book_02}.  In the zero temperature limit (often a suitable approximation in typical experiments which operate at much less than the critical temperature for condensation), the system can be parameterized by a single mean-field ``macroscopic wavefunction" $\Psi({\bf r},t)$.  This encapsulates both the atomic density distribution $n({\bf r},t)=|\Psi({\bf r},t)|^2$ and the coherent phase of the condensate $S({\bf r},t)={\rm arg}[\Psi({\bf r},t)]$.  The spatial and temporal behaviour of $\Psi$ is governed by the Gross-Pitaevskii equation (GPE):

\[i\hbar \frac{\partial \Psi}{\partial t} = \left(- \frac{\hbar^2}{2m} \nabla ^2  + V({\bf r},t) + \frac{4 \pi \hbar^2 a_{\rm s}}{m} |\Psi|^2\right) \Psi,
\]
where $a_{\rm s}$ is the {\it s}-wave scattering length which parameterizes the dominant elastic two-body scattering in the condensate.  The potential $V$ is typically harmonic, e.g. $V({\bf r})=\frac{1}{2}m \left[ \omega_r^2 (x^2+y^2)+\omega_z^2 z^2\right]$ for an axi-symmetric system.  Taking $\omega_z \gg \omega_r$ causes the condensate to become flattened in $z$ and provides access to two-dimensional physics \cite{gorlitz_2001}. 

The hydrodynamical interpretation of $\Psi$ is completed with the definition of a condensate fluid velocity $v({\bf r},t) = (\hbar/m)\nabla S({\bf r},t)$; indeed, within this picture, the GPE corresponds to a classical continuity equation and the Euler equation for an inviscid compressible fluid.  The compressible nature of the condensate is a key difference from Helium, and has important consequences for the turbulent dynamics.

The GPE has proven an accurate description of many aspects of the condensate, from its shape and collective modes to  vortices and phonons \cite{dalfovo_1999,kevrekidis_frantzeskakis_book_08}.  Departures from the GPE become significant, however, at raised temperatures, and a description of finite-temperature extensions to the GPE can be found elsewhere \cite{proukakis_2008,blakie_2008,proukakis_2013,berloff_2014}.

In liquid Helium, the assumption of a local effective interaction is not valid, and while the GPE may provide some qualitative insight of vortex dynamics, it is not physically accurate. A more common model is to treat the vortices as filaments whose mutually-induced dynamics are governed by a Biot-Savart law \cite{barenghi_2014}.  Alternatively, incorporation of a nonlocal interaction term into the GPE provides a closer physical model of Helium \cite{berloff_1999}.

\section{Quantized vortices}
\label{sec:vortices}

Vortices are the ``sinews and muscles of fluid motion" \cite{kuchemann_1965}, and quantum turbulence is dominated by a distribution of quantized vortex lines. The quantization of circulation is a consequence of the coherent phase $S({\bf r},t)$ of the condensate.  To preserve the single-valuedness of $\Psi$, the change in $S$ around any closed path must be quantized as $2\pi q$, where $q=0, \pm 1, \pm 2, \dots$.  Since a gradient in $S$ is related to the fluid velocity, it immediately follows that the circulation $\Gamma=\oint_C {\bf v} \cdot  {\rm d}{\bf l}$ is quantized as $q(h/m)$.  A vortex occurs for non-zero $q$.  Note that $|q|>1$ vortices are energetically unstable compared to multiple singly-charged vortices, and rarely arise unless engineered \cite{leanhardt_2002}.

\begin{figure}[h]
\includegraphics[width=0.5\textwidth]{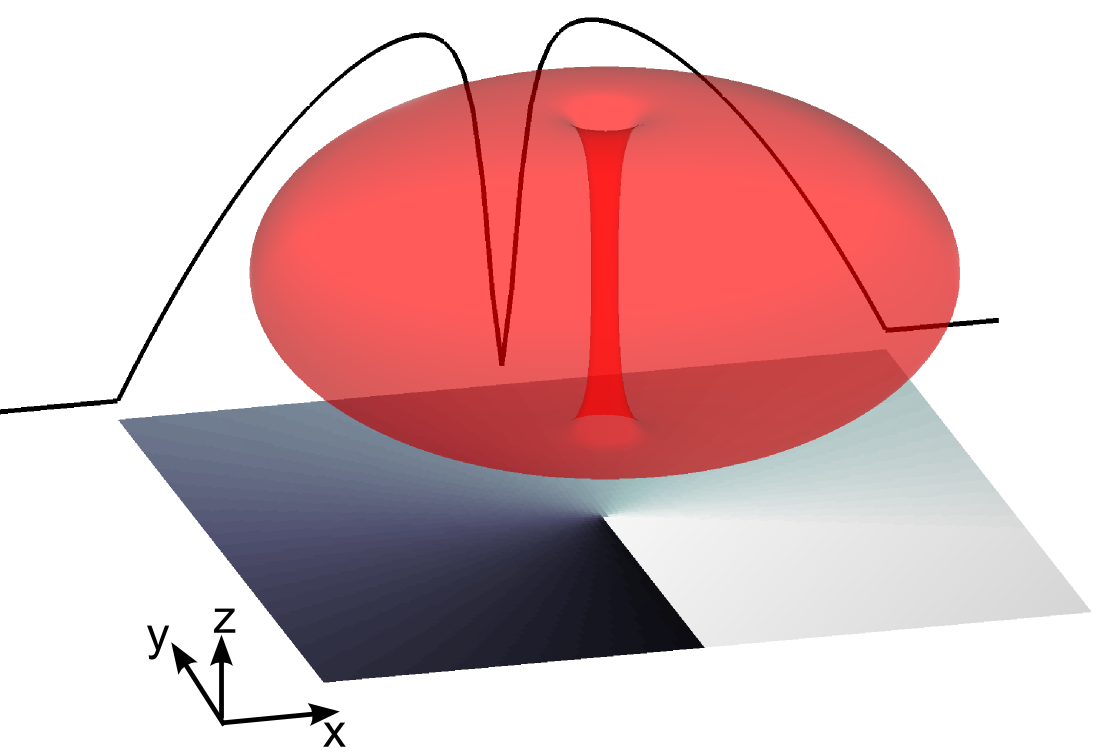}
\caption{Three-dimensional density (isosurface plot) of a trapped condensate featuring a vortex line along the $z$-axis.  The corresponding 2D phase profile and central 1D density profile are also depicted.  }
\label{fig:parker_fig2}
\end{figure}

A schematic of a straight vortex line through a harmonically-trapped condensate is shown in Fig.  \ref{fig:parker_fig2}.  The vortex has a well-defined core, with zero density and the phase singularity at its centre, relaxing to the background condensate density over lengthscale of the order of the healing length $\xi=1/\sqrt{4\pi n a_s}$; this is typically of the order of $0.1-1\mu$m but can be tuned by means of a Feshbach resonance (note that in Helium $\xi \approx 10^{-10}$m is fixed by nature). The condensate circulates azimuthally around the singularity with radial speed $v(r)=q \hbar / (m r)$.   In contrast, in classical fluids (e.g. air, water), vorticity is a continuous field, circulation is unconstrained, and, in the case of filamentary structures (e.g. tornadoes), the core size is arbitrary.

In 3D the vortices may bend, e.g. into tangles and rings, carry helical Kelvin wave excitations and undergo reconnections.  However, under strong axial confinement of the condensate, the dynamics become effectively two-dimensional; here the vortices approach the paradigm of point vortices \cite{middelkamp_2011,aref_2007}.   Being topological defects, vortices can only disappear via annihilation with an oppositely-charged vortex or by exitting the fluid (at a boundary).  The latter effect is promoted by thermal dissipation, which causes a vortex to spiral out of a trapped condensate \cite{jackson_2009,rooney_2010,allen_2013,gautum_2014}.  The role of quantum fluctuations has also been considered \cite{thompson_2012}.

Vortical structures have been generated in the form  of single vortices \cite{madison_2000,freilich_2010}, vortex-antivortex pairs \cite{neely_2010,kwon_2015b}, vortex rings \cite{anderson_2001} and vortex lattices \cite{hodby_2001,aboshaeer_2001}, as well as the disordered vortex distributions of interest here \cite{henn_2009,neely_2013,kwon_2014}.  Methods to generate vortices in condensates include optical imprinting of the phase \cite{leanhardt_2002}, by a rapid quench through the transition temperature for the onset of Bose-Einstein condensation
(i.e., the Kibble-Zurek mechanism \cite{kibble_1976,zurek_1985}) \cite{weiler_2008,freilich_2010}, dragging of an obstacle \cite{raman_2001,neely_2010,kwon_2015a} and mechanical rotation \cite{madison_2000,hodby_2001,aboshaeer_2001}. 

Optical absorption imaging of the vortices is typically preceded by expansion of the cloud to enlarge the cores  \cite{madison_2000,raman_2001}.  This method has been extended to provide real-time imaging of vortex dynamics \cite{freilich_2010}.  While this imaging approach detects density only, the vortex circulation is detectable via a gyroscopic technique \cite{powis_2014}. 


\section{Characteristics}
\label{sec:nature}

\subsection{Scales}
Under suitable continuous forcing and dissipation, turbulence reaches a statistical steady state.  This is characterised by being statistically self-similar in time and space, with physical quantities behaving according to scaling laws.  In ordinary turbulence these ranges can be vast, e.g. the lengthscale of eddies in atmospheric turbulence span the range $10^{-4}-10^{4}$~m.  In contrast, the accessible lengthscales in condensates are limited from approximately $10^{-6}$~m (the healing length $\xi$; smallest scale of density variations) to $10^{-4}$~m (the system size $D$), and the true range of self-similarity may be considerably narrower.  This raises questions as to the nature of turbulence on such restricted scales, including whether it is justifiably ``turbulent".  Nonetheless, condensates provide a useful setting for analysing vortex dynamics, from chaos and self-ordering to the quantum-classical crossover.  Of great benefit is that all relevant fluidic scales in the system can be simulated in unison, something not possible for many classical scenarios.  

Large-scale flow in quantum turbulence occurs through coherent vortex structures. For example, in 2D, clusters of like-sign vortices form a collective macroscopic flow which, when coarse-grained, mimics classical vorticity.  In 3D the equivalent is for localized bundling of like-sign vortex lines \cite{baggaley_2012b}, as depicted in Fig. \ref{fig:parker_fig1}(b).  These structures provide the intuition of approaching classical behavior as the number of quanta is increased.  

Another important lengthscale is the typical inter-vortex distance $\delta$.  In condensates the lengthscales $\xi$, $\delta$ and $D$ are compact, with roughly one order of magnitude separating them [Fig.~\ref{fig:parker_fig1}(a)].   This should be compared with the many orders of magnitude which separate these distances in Helium II [Fig.~\ref{fig:parker_fig1}(b)]; it is a result of this that vortex filament models are the workhorse for modelling turbulence therein, rather than the microscopic GPE.  The interactions of vortices with each other and with the boundaries are magnified, which in turn promotes the occurence of processes such as reconnections and nucleation/loss of vortices at the boundary.

\subsection{3D quantum turbulence}
Quantum turbulence is two-faced: on one side sharing common properties with ordinary turbulence; on another showing striking non-classical phenomena.  Steady-state turbulence in a bulk ordinary fluid, forced at some large-scale, is characterised by the famous Kolmogorov behaviour: energy is transferred from larger to smaller scales without dissipation over an inertial range of wavenumber $k$, leading to the Kolmogorov spectrum $k^{-5/3}$ of energy.  This process is called the Richardson cascade: large-scale eddies, created at the forcing lengthscale, evolve into progressively smaller eddies.  Remarkably, this classical Kolmogorov spectrum emerges in vortex filament calculations \cite{barenghi_2014} which model Helium, and GPE simulations of 3D quantum turbulence \cite{nore_1997,kobayashi_2005,yepez_2009} in the spectral range $k \gg 2 \pi / \xi$.  This classical-like behaviour is believed to arise from bundling of vortex lines \cite{baggaley_2012b,barenghi_2014}, that is, vortex lines of the same polarity come together, forming metastable structures.  Note that, to link to classical incompressible fluids, the relevant energy in the compressible quantum fluid is the incompressible kinetic energy, i.e. the energy associated with the vortices (the remainder of the kinetic energy, the compressible part, is associated with phonons) \cite{white_2014}.   Reconnections between vortex lines are believed to facilitate this cascade by producing small vortex rings.  The vortex lines support helical Kelvin wave excitations, which become excited during reconnection events.  These excitations carry energy at relatively small scales (high $k$) and lead to a $k^{-3}$ scaling of the incompressible kinetic energy in the range $k \lappeq 2\pi /\xi$ \cite{yepez_2009}.  These results were obtained in a large homogeneous system, and it remains to be established if this scaling behaviour persists in trapped condensates and whether this can be experimentally detected.

Further differences emerge in the turbulent velocity field.  In classical isotropic turbulence, the probability distribution of the fluid velocity components is a near-Gaussian.   However, in quantum turbulence the velocity statistics have a power-law behaviour, as confirmed experimentally in Helium \cite{paoletti_2008} and through GPE-based simulations of trapped condensates \cite{white_2010}.  This distinction arises from the singular nature of quantized vorticity and the $1/r$ velocity profile.  However, classical near-Gaussian behaviour is recovered by coarse-graining over the typical intervortex distance \cite{baggaley_2011,mantia_2014}.

Due to their compressible nature, condensates also support self-similar cascades in acoustic energy, which can be analysed in the context of weak-wave turbulence \cite{proment_2009}.

The typical harmonic trapping of condensates leads to a background density which is inhomogeneous throughout and soft boundaries.  The recent achievement of an atomic condensate in a box-like trap \cite{gaunt_2013} allows for the study of quasi-homogeneous turbulence and thus the crossover towards the homogeneous setting of Helium.

\subsection{2D quantum turbulence}
Highly-flattened condensates provide access to two-dimensional quantum turbulence.  As known from classical fluids, 2D turbulence is vastly different to its 3D counterpart.  The vortices become rectilinear, negating Kelvin waves, and reconnections become annihilation events.  More importantly, however, 2D incompressible fluids possess an additional conserved quantity, the ``enstrophy", which is the total squared vorticity \cite{kraichnan_1980,boffetta_2012}.  This has profound consequences: from some forcing lengthscale, energy is transferred to {\it larger} scales, associated with the aggregation of vorticity into larger structures.  Jupiter's Great Red Spot is the natural exemplar of this {\it inverse cascade}.  The inverse cascade leads to a $k^{-5/3}$ scaling of incompressible kinetic energy for $k < k_{\rm f}$ (where $k_{\rm f}$ is the forcing wavenumber); additionally, enstrophy is transported towards smaller scales, leading to a $k^{-3}$ scaling for $k > k_{\rm f}$.  

In the quantum fluid, enstrophy is simply proportional to the number of vortices.  However, compressibility allows for vortices of opposite sign to annihilate, breaking the conservation of enstrophy, and also introducing interactions with phonons.  Nonetheless, Reeves {\it et al.} \cite{reeves_2013} were able to suppress these compressibility effects (by suitable forcing and dissipation) and demonstrate the inverse cascade within GPE simulations.  This arose through the clustering of vortices with like-sign, and progressive increasing in their net charge and spatial scale (thereby mimicking rigid-body rotation \cite{reeves_2014}).   The incompressible kinetic energy was confirmed to follow a $k^{-5/3}$ scaling, with accumulation at the largest scales of the system. 

The role of the vortices in the condensate's kinetic energy spectrum has been studied theoretically \cite{bradley_2012}.  For $k > \xi^{-1}$ the spectrum is $k^{-3}$ and associated with the vortex core structure, while for $ k< \xi^{-1}$ the spectrum follows the $k^{-5/3}$ which is determined by the distribution of the vortices.  Furthermore, the system allows one to study the crossover to weak-wave turbulence \cite{nazarenko_2006,reeves_2012}.

\subsection{Vortex-sound interactions and the acoustic sink} 

In ordinary fluids, the sink at small-scales is provided by viscosity, which dissipates small-scale eddies into heat.  In finite temperature quantum fluids, an effective viscosity due to the interaction of the condensate with the thermal cloud provides a similar dissipation \cite{berloff_2014}.  In the zero temperature limit, however, viscosity is absent; then dissipation is thought to occur at small lengthscales through the decay of vortical motion into phonon excitations (sound waves) \cite{barenghi_2005}.  

\begin{figure}[h]
\includegraphics[width=0.44\textwidth]{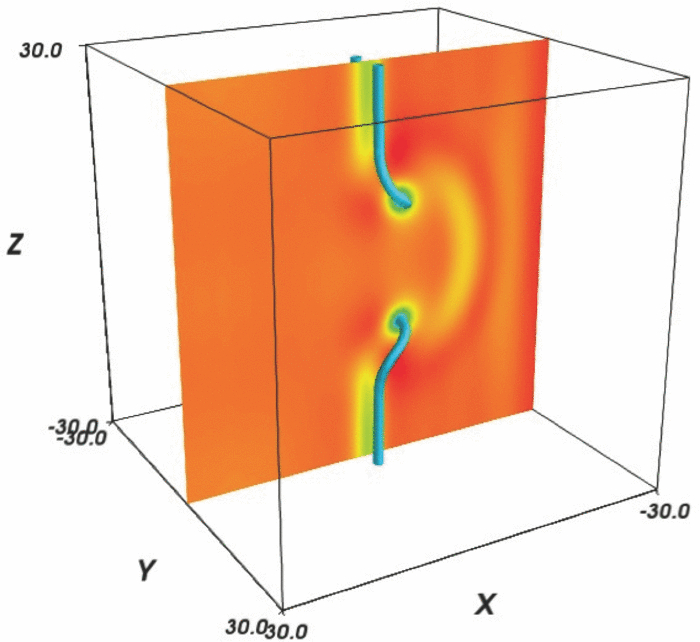}~~~~~
\includegraphics[width=0.38\textwidth]{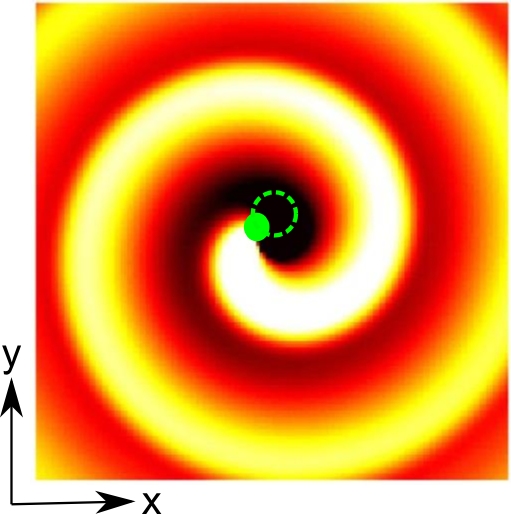}
\caption{(a) Sound emission from a reconnection of two vortex lines (blue).  Figure taken from \cite{zuccher_2012}.  (b) A vortex moving in a circular path radiates sound waves in a quadrupolar pattern, distorted into a spiral by the vortex motion \cite{parker_2004}.}
\label{fig:parker_fig3}
\end{figure}  
One such contribution comes from vortex reconnections, where the abrupt change in the fluid topology and velocity field produces a burst of sound waves, as predicted in GPE simulations \cite{leadbeater_2001,zuccher_2012} and depicted in Fig.~\ref{fig:parker_fig3}(a).  The sudden snap-like dynamics of reconnections is so dominant that temperature has no significant effect \cite{allen_2014b,bewley_2008}.  In 2D, the analog is the sound-generating annihilation of a vortex-antivortex pair.  Another contribution occurs under acceleration of a vortex line segment, which lead to the emission of sound waves \cite{vinen_2001,parker_2004}, depicted in Fig. \ref{fig:parker_fig3}(b) for a precessing 2D vortex.  This is analogous to the electromagnetic (Larmor) radiation from an accelerating charge \cite{arovas_1997}.  However, this transfer of incompressible, vortex energy to compressible, sound energy is not one way: sound energy can be re-absorbed by the emitting vortex or another vortex \cite{parker_2004,parker_2012}.  Evidence suggests that these sound-mediated interactions should not disrupt the inverse cascade in 2D quantum turbulence \cite{lucas_2014}.

%
%

\section{Experiments}
\label{sec:experiments}

\subsection{3D quantum turbulence (Sao Paulo)}
In 2009 Henn {\it et al.} \cite{henn_2009} reported the generation of quantum turbulence in an elongated condensate via an oscillating trap perturbation.  While weak agitation created collective modes of the condensate, i.e. dipole, quadrupole and scissors modes, stronger agitation led to vortices becoming nucleated into the condensate (likely through a dynamical instability of the condensate surface, analogous to that arising under rotation \cite{parker_2006}).   Initially well-separated, the vortices increased in density with agitation until a disordered tangle of vortex lines permeated the condensate  [Fig. \ref{fig:parker_fig3}].  Under 2D absorption images the vortex tangle has low contrast, and it remains a challenge to image the 3D vortical structure, although recent experimental advances in simultaneously imaging in two directions may assist in vortex identification \cite{lamporesi_2013}.

\begin{figure}[h]
\includegraphics[width=0.7\textwidth]{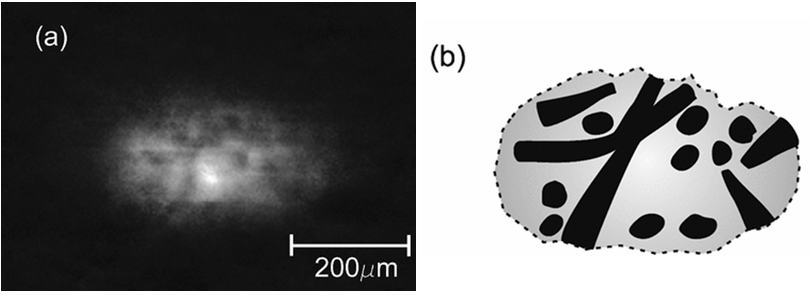}
\caption{(a) Absorption image of the condensate (following expansion) reported by Henn {\it et al.} \cite{henn_2009}.  The dark filamentary regions represent vortex lines. (b)  An illustration of the inferred vortex distribution in (a).  Figures taken from Ref. \cite{henn_2009}. }
\label{fig:parker_fig4}
\end{figure}  

Following release from the trap, the condensate expanded with approximately constant aspect ratio, in contrast to the inversion of aspect ratio expected for ordinary condensates \cite{pethick_smith_book_02}.   This anomalous expansion is consistent with the distribution of vorticity throughout the cloud  \cite{caracanhas_2012}, and thus provides an experimental signature of isotropic quantum turbulence. Later work has shown that the turbulent and non-turbulent condensates have very different momentum distributions \cite{thompson_2014}.

\subsection{2D quantum turbulence (Arizona)}

Neely {\it et al.} \cite{neely_2013} employed a flattened condensate with harmonic trapping ($\omega_z/\omega_r \sim 10$), pierced in the centre by a localized repulsive Gaussian obstacle from a blue-detuned laser beam so as to form an annular net potential.  The relative motion of an obstacle through a condensate is known to generate vortex-antivortex pairs when the motion exceeds a critical speed \cite{frisch_1992,stagg_2015b}.  Keeping the obstacle fixed, the harmonic trap was translated off-centre and moved in a circular path for a short time, before returning to its original position.  This generated a disordered distribution of approximately 20 vortices [Fig. \ref{fig:parker_fig4}].  Note the high visibility of the vortices, due to their alignment along $z$ as per the 2D geometry.  The number of vortices decayed over time, until reaching a final state of a (vortex-free) persistent current in the annular trap.

\begin{figure}[h]
\includegraphics[width=1\textwidth]{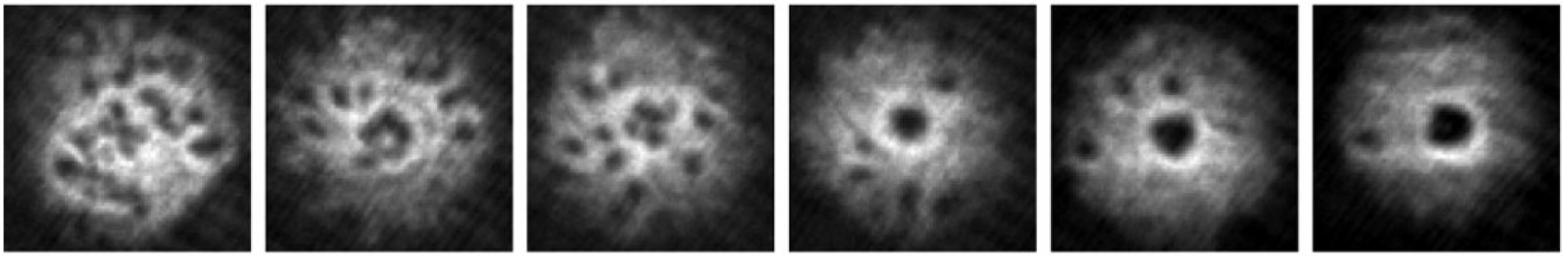}
\caption{Absorption images of the condensate in \cite{neely_2013} at various times following cessation of the stirring (from left to right: 0, 0.15, 0.33, 0.67, 1.17 and 8.17 s).  An initial disordered vortex distribution evolves towards a persistent current.  Figure taken from \cite{neely_2013}.}
\label{fig:parker_fig5}
\end{figure}

These dynamics bear analogy to the inverse cascade of isotropic 2D compressible turbulence: small-scale stirring generated small-scale vortical excitations, which self-order over time into a large-scale flow (the persistent current).  However, the stirring imparts angular momentum to the condensate, which may drive the evolution towards the persistent current.   

Numerical simulations of the experiment reveal that stirring leads to the formation of a Kolmogorov-like $k^{-5/3}$ spectrum of the incompressible kinetic energy for $k < k_s$ and a $k^{-3}$ spectrum for $k > \xi^{-1}$.  Additionally, vortices of the same sign tend to form clusters.  These features becomes gradually lost once the stirring ceases, as vortices decay towards the persistent current.

\subsection{2D quantum turbulence (Seoul)}
Kwon {\it et al.} \cite{kwon_2014} employed a similar set-up as above, apart from the harmonic trap begin translated linearly (in $x$) relative to the obstacle and subsequently removed.  When the translation speed $v$ exceeded a critical value ($\sim 0.5$ mm s$^{-1}$), vortices were nucleated into the condensate \cite{kwon_2015a,kwon_2015b}.  The number of vortices $N_{\rm v}$ increased with $v$, saturating at around 60 vortices [Fig. \ref{fig:parker_fig5}(a)].   The vortices were distributed in a disordered manner, characteristic of 2D quantum turbulence [Fig. \ref{fig:parker_fig5}(b)], and decayed over time.

\begin{figure}[h]
\includegraphics[width=0.95\linewidth]{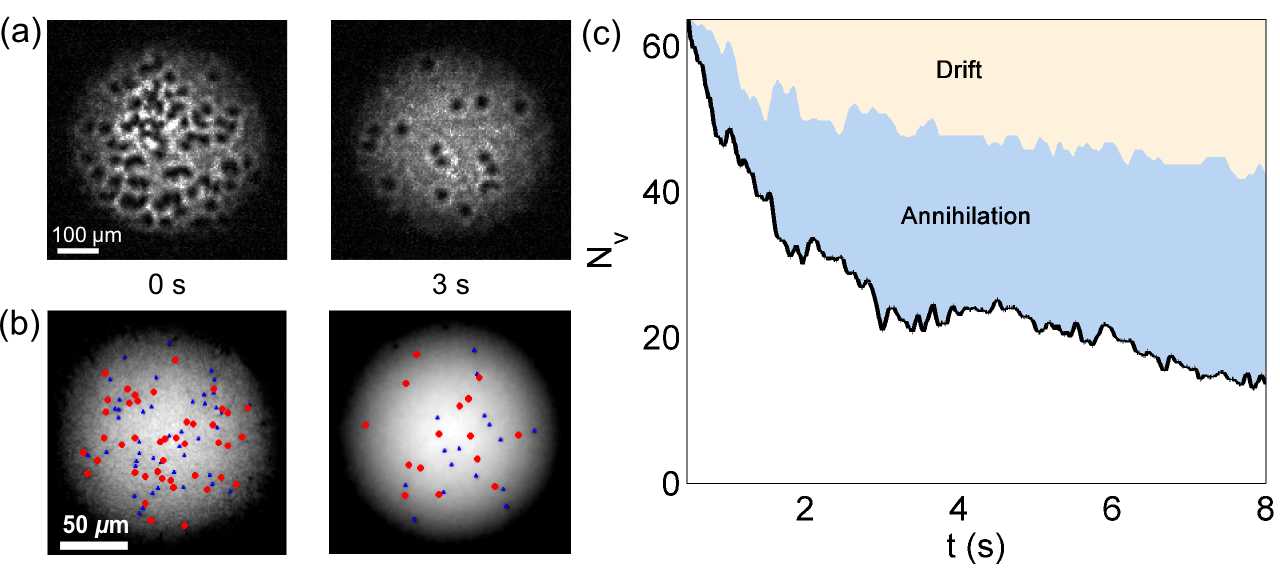}
\caption{(a) Absorption images of the expanded condensate \cite{kwon_2014} at various times (since removal of the obstacle).  Images courtesy of Yong-il Shin. (b) Corresponding images of (unexpanded) condensate density from dissipative GPE simulations \cite{stagg_2015a}.  Vortices with positive (negative) circulation are highlighted by red circles (blue triangles).  The vortices appear much smaller since the condensate has not been expanded.  (c) Decay of the vortex number $N_{\rm v}$, with the contribution of drifting and annihilation depicted by the shaded regions.
}
\label{fig:parker_fig6}
\end{figure}  

The experiment focussed on this decay of vorticity and suggested two contributions: (i) thermal dissipation (resulting in drifting of vortices to the edge of the condensate), and (ii)  vortex-antivortex annihilations.  To support this picture, it was found that the decay of $N_v$ followed the form:
\begin{equation}
\frac{{\rm d}N_{\rm v}}{{\rm d}t} = - \Gamma_1 N_{\rm v} - \Gamma_2 N_{\rm v}^2,
\label{eqn:parker_eqn2}
\end{equation}
where the linear and quadratic terms, parameterized by the coefficients $\Gamma_1$ and $\Gamma_2$, model these two decay processes.  The decay was examined at various temperatures, with both $\Gamma_1$ and $\Gamma_2$ increasing with temperature.  Crescent-shaped density-depleted regions occasionally appeared in the images, and were associated with vortex-antivortex annihilation events.

Numerical simulations \cite{stagg_2015a} have shed further light on these dynamics.  As the condensate moves relative to the obstacle, vortices are nucleated, often in clusters of like-sign, forming quasi-classical wakes.  The condensate sloshes in the trap, generating further vortices and mixing the new with the old.  Over time the clusters disaggregate and the vortices become randomized.  Here the condensate is characterised by a disordered distribution of vortices (with no net circulation), collective modes and
a tempestuous field of sound waves, indicative of two-dimensional quantum turbulence.  Incorporation of phenomenological dissipation into the simulations (to mimic thermal dissipation) leads to a decay of $N_{\rm v}$ consistent with Eq. (\ref{eqn:parker_eqn2}).  The separate decay contributions from drifting and annihilations can be identified, and show that as $T \rightarrow 0$ drifting becomes negligible relative to annihilation.

\section{Generation}
\label{sec:generation}

The particular method used to generate turbulence can strongly bias the turbulent state which forms, and a key challenge is to establish experimentally-feasible approaches which promote isotropic and homogeneous turbulence in condensates, despite their small size (relative to the vortex core size), soft boundaries and sensitivity to heating.  In turn, the generation method may allow control over the nature of the turbulence, e.g. the forcing scale, the transition from chaotic dynamics of few vortices to turbulence, and the ratio of compressible to incompressible kinetic energy.   The experimental stirring methods above tend to impart momentum (linear and/or angular) to the condensate and excite collective modes.  Simple modifications of these strategies may promote isotropy: stirring in a figure-of-eight path (rather than circular) leads to negligible momentum transfer and an isotropic velocity field \cite{allen_2014}, while making the Gaussian obstacle elliptical enhances vortex nucleation, which allows for reduced stirring speeds and in turn reduced surface modes and phonons \cite{stagg_2014,stagg_2015a,stagg_2015b}.

While rotation of a condensate (in an anisotropic trap) leads to vortex lattices, Kobayashi {\it et al.} \cite{kobayashi_2007} showed that rotation about a second axis leads to a dense tangle of vortices with Kolmogorov behaviour.  An attractive feature of this set-up is that it provides access to the crossover from isotropic turbulence to polarized turbulence to vortex lattices.  

Vortices may be imposed deterministically as an initial state by phase imprinting \cite{leanhardt_2002}, and a recent variation of this, based on orbital angular momentum transfer from a holographic light beam \cite{brachmann_2011} allows for creation of vortex configurations with almost arbitrary charge and geometry.  What's more, relatively simple vortex structures initially present in the condensate may undergo an instability to produce quantum turbulence \cite{horng_2008,horng_2009}.  A turbulence vortex tangle is also predicted to occur during the evolution of a Bose gas fron highly non-equilibrium conditions \cite{berloff_2002}.

While turbulence is associated with no net vorticity, it is possible to create disordered distributions of vortices dominated by one sign of circulation, e.g. as transient states in rotating condensates \cite{schweikhard_2004,parker_2005,wright_2008}, through stirring \cite{white_2012} and the decay of a giant vortex \cite{santos_2014}. Although not strictly turbulent, these systems offer insight into vortex clustering and self-organization processes.

The generation of statistical steady states of quantum turbulence, with suitable forcing and dissipation, is more challenging in trapped condensates.  However, a significant step has been achieved numerically \cite{reeves_2013}: with forcing provided by flow past a series of obstacle and phenomenological damping, a steady state of 2D quantum turbulence was achieved.  
Recent simulations have also shown that vortex nucleation and the formation of vortex tangles in three-dimensions may be enhanced at finite temperature \cite{stagg_2015c}.

\section{Exotica}
\label{sec:exotica}

Condensates offer access to exotic forms of quantum turbulence; here we list some topical examples.  

\begin{figure}[h]
  \includegraphics[width=0.62\textwidth]{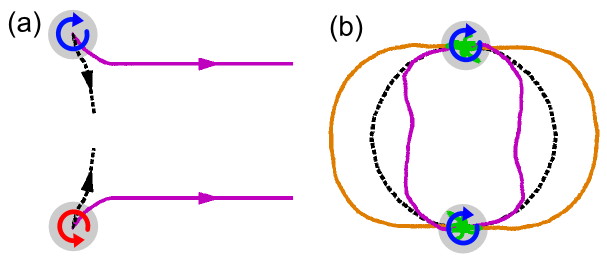}~~~~
   \includegraphics[width=0.27\textwidth]{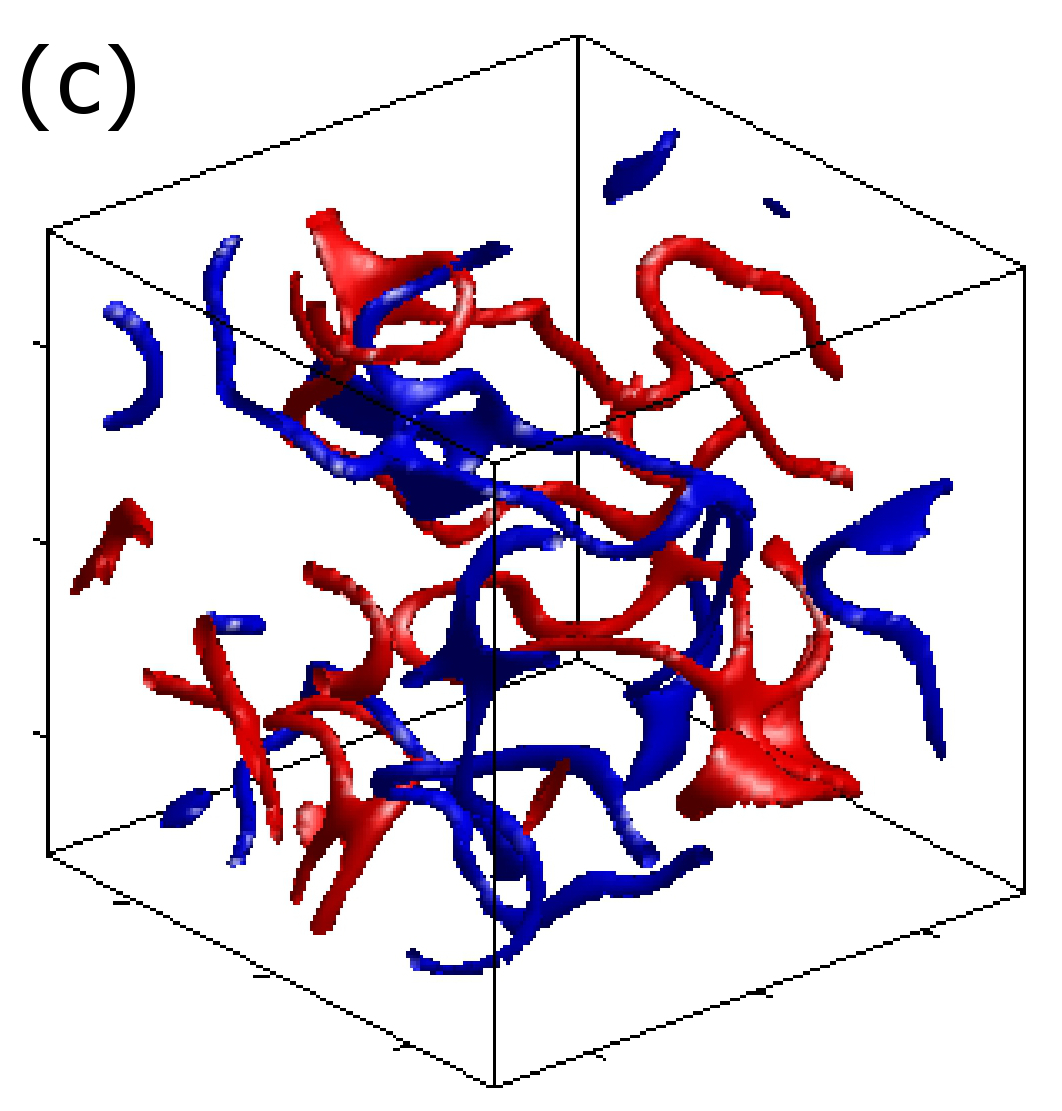}
 \caption{(a, b) Vortex dynamics in a quantum ferrofluid \cite{mulkerin_2013}.  Dipolar interactions can (a) stabilize vortex-antivortex pairs against annihilation and (b) induce anisotropic corotation for vortex-vortex pairs.  The non-dipolar dynamics are shown by black dotted lines. (c) Turbulence in a binary condensate can exist as a two inter-coupled vortex tangles, shown here through a simulation of the binary GPE \cite{pattinson_2015}.}\label{fig:parker_fig7}
\end{figure}

\paragraph{Quantum ferrofluid turbulence:}
Condensates have been formed with $^{52}$Cr \cite{griesmaier_2005}, $^{164}$Dy \cite{lu_2011} and $^{168}$Er \cite{aikawa_2012} atoms.  These possess significant magnetic dipole moments, such that the atomic interactions develop a long-range and anisotropic dipolar contribution (in contrast to the short-range and isotropic van der Waals interactions).  The quantum fluid then acquires a global ferromagnetic behaviour, forming a ``quantum ferrofluid" \cite{lahaye_2009}.  Each vortex then behaves as a macroscopic dipole, introducing an additional interaction between vortices which is long-range and anisotropic \cite{mulkerin_2013,mulkerin_2014}.  This has significant consequences for the vortex dynamics, controlling the annihilation threshold for vortex-antivortex pairs and the anisotropy of the vortex-vortex dynamics [Fig. \ref{fig:parker_fig7} (a, b)].  This can be expected to lead to rich behaviour in the ensuing turbulence, and provide a handle on the role of annihilations.

\paragraph{Onsager vortices and negative temperatures:}

In 1949 Onsager considered turbulence in 2D classical fluids as a disordered collection of point vortices \cite{eyink_2006}.  This model has a finite phase space, the consequence of which is that as energy is added the system it will become more ordered, i.e. lose entropy, a process associated with negative temperature.  Remarkably, this phenomenon appears accessible in compressible 2D condensates \cite{billam_2014,simula_2014}.  Out of an initially disordered distribution of $\pm$ vortices in an isolated (undriven, undamped) system, giant long-lived macroscopic clusters of like-signed vortices emerge \cite{simula_2014}, the ``Onsager vortices".  This self-ordering process is associated with the evaporation of vortices, which increases the average energy per vortex, a so-called ``evaporative heating".  This provides a setting to understand a universal phenomena of vortex clustering and the counter-intuitive realm of negative temperatures.   

\paragraph{Multicomponent turbulence:} Several co-existing condensates can be formed, of either different atoms, different isotopes or different hyperfine states \cite{kevrekidis_frantzeskakis_book_08}.  The components are coupled, and may either be miscible or immiscible, depending on a play-off between the inter-species and intra-species interactions; in this manner, turbulence may exist as a multiple interpenetrating and coupled vortex tangles, as depicted for a two-component (binary) condensate in Fig. \ref{fig:parker_fig7}(c).    

The presence of inter-component hydrodynamical instabilities opens up new mechanisms to generate quantum turbulence.  For miscible binary condensates made to counter-propagate, a counterflow instability develops above a critical relative speed; this leads to vortex nucleation and ultimately two interpenetrating turbulent quantum fluids \cite{takeuchi_2010}.   This is analogous to the thermal counterflow method of $^4$He; however, vortices are generated intrinsically within the bulk rather than through ``remnant" vortices.  For immiscible binary condensates \cite{kobyakov_2014}, a Rayleigh-Taylor instability can set in at the interfacial boundary, leading to mixing of the condensates; a Kelvin-Helmholtz instability subsequently produces vortex bundles, which can evolve into turbulence.  Unlike other methods for generating quantum turbulence, this can take place at arbitrary small Mach number.  Aspects of universal behaviour in the dynamics of binary condensates far from equilibrium have also been considered \cite{karl_2013}.

Multicomponent condensates may possess spin degrees of freedom, termed ``spinor" condensates, which allow for unconventional topological defects, including fractional vortices, monopoles and skyrmions, and rich superfluid dynamics \cite{kasamatsu_2005,stamper_2013}.  In particular, it is possible to generate spin turbulence, in both homogeneous and trapped condensates, whereby the spin density vectors are disordered \cite{fujimoto_2012,fujimoto_2014}.  Spin turbulence obeys characterstic scaling laws which are distinct from conventional quantum turbulence, and depends on the form of the spin-dependent interaction.  The spin configuration is also coupled to the density, and so coupled spin-density turbulence is further possibility for unconventional turbulence.

\section{Outlook}
\label{sec:conclusions}
The study of turbulence in quantum gases is still in its infancy.  The discrete nature of the vorticity and the scarcity of dissipation (particularly in the low temperatue limit $T \ll T_{\rm c}$, where $T_{\rm c}$ is the critical temperature) make quantum fluids ideal testing grounds to explore the emergence of classical behaviour observed in ordinary fluids, such as the Kolmogorov spectrum \cite{baggaley_2012b, maurer_1998,salort_2010} and Gaussian velocity statistics \cite{baggaley_2011,mantia_2014}, from many elementary quanta of circulation.  In other words, under certain conditions quantum turbulence seems to capture the main properties of turbulence, thus representing, perhaps its skeleton.  On the other hand, the study of this new form of turbulence has also revealed new physics which is worth studying per se.  The study of quantum turbulence in atomic BECs presents opportunities (such as the transition from 2D to 3D turbulence) as well as theoretical and experimental challenges (such as the finite-temperature generalization of the GPE and the visualization of 3D turbulence).
\bibliography{turbulence_refs}
\bibliographystyle{cambridgeauthordate}

\end{document}